\documentclass[pra, superscriptaddress, twocolumn]{revtex4}

\usepackage{empheq}
\usepackage{dcolumn}
\usepackage{amsmath,amssymb}
\usepackage{bm}
\usepackage{color}
\usepackage{overpic}
\usepackage{latexsym}
\usepackage{color}
\usepackage{graphicx}
\usepackage{epstopdf}
\usepackage{psfrag} 
\usepackage{epsf} 
\usepackage{subfigure} \usepackage{amsfonts}
\usepackage{bm}
\usepackage{appendix}
\usepackage{hyperref}
\usepackage{comment}

\def\beq{\begin{equation}}
\def\eeq{\end{equation}}
\def\ba{\begin{align}}
\def\enda{\end{align}}
\def\bi{\begin{itemize}}
\def\ei{\end{itemize}}

 \newcommand{\ket}[1]{|#1\rangle}
 
 \newcommand{\braket}[2]{\langle #1|#2\rangle}

\begin{document}

\title{Ground state of composite bosons in low-dimensional graphs}
\author{Cecilia Cormick}
\affiliation{Instituto de F\'{i}sica Enrique Gaviola, CONICET and Universidad 
Nacional de C\'{o}rdoba,
Ciudad Universitaria, X5016LAE, C\'{o}rdoba, Argentina}
\author{Leonardo Ermann}
\affiliation{Departamento de F\'isica Te\'orica, GAIDI, Comisi\'on Nacional de Energ\'ia At\'omica, Buenos Aires, Argentina}
\affiliation{CONICET, Godoy Cruz 2290 (C1425FQB) CABA, Argentina}
\affiliation{ECyT-UNSAM, Campus Miguelete, 25 de Mayo y Francia, 1650 Buenos Aires, Argentina}

\date{April 28, 2023}

\begin{abstract}
We consider a system of composite bosons given by strongly bound fermion pairs tunneling through sites that form a low-dimensional network. It has been shown that the ground state of this system can have condensate-like properties in the very dilute regime for two-dimensional lattices but displays fermionization for one-dimensional lattices. Studying graphs with fractal dimensions, we explore intermediate situations between these two cases and observe a correlation between increasing dimension and increasing condensate-like character. However, this is only the case for graphs for which the average path length grows with power smaller than 1 in the number of sites, and which have an unbounded circuit rank. We thus conjecture that these two conditions are relevant for condensation of composite bosons in arbitrary networks, and should be considered jointly with the well-established criterion of high entanglement between constituents.
\end{abstract}

\maketitle

\section{Introduction} \label{sec:intro}

Several recent developments have opened the doors to the exploration of systems with very complex geometries, approaching the freedom to build arbitrary networks for many-body physics. This is possible due to the engineering capabilities achieved with optical tweezer arrays \cite{Endres_2016, Schymik_2020} and most notably with circuit quantum electrodynamics \cite{Kollar_2019, Carusotto_2020}. Other examples in this direction are given by the implementation of frustrated lattices and quasicrystalline structures in optical traps \cite{Jo_2012, Sbroscia_2020}, photonic quasicrystals \cite{Freedman_2006} and polaritonic systems \cite{Baboux_2017}, as well as artificial fractal lattices for electrons \cite{Kempkes_2018} and tree-like couplings of atom arrays inside optical cavities \cite{Periwal_2021}.
Implementations in the context of Josephson-junction arrays have also inspired the study of Bose-Hubbard models on various graphs \cite{Burioni_2000, Buonsante_2004, Sodano_2006}.

It is well known that the dimensionality of a physical system can have tremendous impact on its properties. One remarkable illustration of this phenomenon is given by the fact that hard-core bosons exhibit fermionization in one-dimensional (1D) systems \cite{Tonks_1936, Girardeau_1960}, thus questioning intuitive expectations of what bosonic behaviour means. Another striking example is the existence of anyonic particles only in two dimensions \cite{Arovas_1985}. In more general scenarios, other graph properties besides dimension have been shown to play a relevant role as well, for instance for the observation of Bose-Einstein condensates in low-dimensional graphs \cite{Burioni_2000}, propagation of solitons in networks \cite{Sobirov_2010}, or frustration in spin systems \cite{Diep_Book}.

The number of spatial dimensions has been recently studied in connection with the validity of the so-called ``coboson ansatz'' for the ground state of a system of $N$ composite bosons \cite{Cespedes_2018} . This ansatz is a canonical-ensemble analogue of the BCS wavefunction and can in many cases be used as a compact approximation of the ground state of a many-particle system \cite{combescot_2008}. We note that the ansatz describes approximate condensation to the ground state of a single composite particle, as opposed to a more general kind of condensation to an arbitrary state as studied in \cite{Tennie_2017}. 

Loosely speaking, the coboson ansatz is expected to be suitable for dilute systems with short-range interactions and high entanglement between the constituents of each composite boson \cite{combescot_2008, tichy_2012b}. It has been noted that dimensionality also plays an important role for the validity of the ansatz, and that 1D systems fulfilling the previous conditions do not generally obey the ansatz \cite{Cespedes_2019, Cuestas_Cormick_2022}. This is not surprising in the light of the already mentioned fermionized behaviour. However, fermion pairs in a ladder geometry, i.e. an $n\times m$ lattice with $n\to\infty$ while $m$ is kept fixed, behave as in a 1D system although fermionization is not applicable \cite{Cespedes_2019}. We note that this dimension-dependent change in behaviour, fermionizing in 1D but condensing in 2D for low enough densities, was first studied for standard hard-core bosons, see for instance \cite{Bernardet_2002}.

We now extend the previous analysis of the relation between lattice dimensionality and validity of the coboson ansatz to models with dimensions between 1 and 2. To this aim, we consider fractal geometries, and compare the performance of the ansatz for fractals with different Hausdorff dimensions \cite{Falconer_2003}. The lack of translational invariance makes this study numerically costly, so that the systems studied have moderate sizes. Nevertheless, several conclusions can be drawn from our results. Indeed, the graph dimension does seem to play an important role for the validity of the ansatz. However, a new aspect that appears in our analysis is the observation that fermion pairs do not exhibit condensation in tree-like graphs. Besides, we find that the ansatz is generally a much better approximation to the ground state in systems with closed boundaries, while the number of neighbours has little impact on the results.

Because of the computational cost, we focus on systems of a few fermion pairs, i.e. two or three hard-core bosons or equivalently four or six paired fermions. The structure of coboson theory is such that the coboson ansatz for an arbitrary number $N$ of pairs leads, for low densities, to expectation values dominated by the results of one and two pairs, accompanied by known $N$-dependent prefactors \cite{combescot_2008}. Then, if the ansatz is shown to give wrong results for two composite bosons, it cannot provide a good approximation of the ground state of $N$ pairs either. 

This article is organized as follows: in Sec. \ref{sec:model} we introduce the model we study for different networks. Section \ref{sec:cobosons} discusses basic concepts regarding the coboson ansatz that are essential to understand our work, whereas Sec.~\ref{sec:graphs} introduces relevant graph properties. In Sec. \ref{sec:results} we present our numerical results for various graphs with dimensions between 1 and 2. Finally, in Sec. \ref{sec:conclusions} we summarize our conclusions.

\section{The model} \label{sec:model}

The system we consider consists of a graph of $M$ sites with two species 
of fermions that can hop along them. Fermions 
of different species experience a very strong attraction, so that if the numbers of particles of
both species are equal then the low energy manifold has all particles in pairs. 
More precisely, the Hamiltonian takes the form:
\begin{multline}
H = -U_0 \sum_{j=1}^M a_j^\dagger a_j \, b_j^\dagger b_j \\+ \frac{J}{2} 
\sum_{<i,j>} (a_i^\dagger a_j +
b_i^\dagger b_j + {\rm H.c.} )
\label{eq:H}
\end{multline}
where $a_j$ ($a_j^\dagger$) destroys (creates) a particle of type $a$ in site 
$j$, $b_j$ ($b_j^\dagger$) does the same for a particle of type $b$, and $<i,j>$ indicates neighbouring sites in the graph. 
For simplicity, we do not consider spatial variations of local energies or tunneling strength.

We are interested in the limit when the interaction energy is much stronger than the hopping, i.e. 
$U_0\gg J$, and apply perturbation theory to find the ground state of the system. 
As will be shown in the following, the restriction to the limit 
when particles always tunnel in pairs makes our system an instance of the 
hard-core Bose-Hubbard model, which is equivalent to a Heisenberg model 
\cite{baxter_book, bethe_1931, karabach_1997, Blundell_book}. Heisenberg models on fractal lattices have been studied for instance in \cite{Tomczak_1996, Voigt_1998, Voigt_2001}. An important point to keep in mind is that for a coboson system with $N$ composite particles only a particular subspace of the equivalent spin system with a fixed total spin projection $S_z$ is relevant \cite{Cespedes_2019}.

This limit of 
very strongly bound pairs is the one studied in \cite{tichy_2012, Cespedes_2019}, and it is particularly 
relevant for our purposes since it is the situation where the coboson description should be most appropriate. The Hilbert space of $N$ fermions of each kind divides into a ground manifold composed by the states where all fermions are paired (i.e. 
occupying the same site as one of the other species), and many excited manifolds with unpaired particles. The 
effective Hamiltonian within the ground manifold can be found with perturbation theory. We note that the steps involved are the same as in \cite{tichy_2012, Cespedes_2019}, but now generalized to arbitrary graph geometries.

To zero order in the hopping the energy of 
the ground manifold is $-NU_0$. When second-order terms in the hopping are introduced, one
obtains an overall shift of the energy of this subspace, Hamiltonian terms describing correlated tunneling of pairs, and an additional term coming from the fact that hopping 
of a particle into a given site is forbidden if there is already an identical fermion 
there. This leads to the form:
\beq
H_g^{(N)} \simeq -N \left(U_0+ J^{\rm eff}\right) + H^t + H^d \,.
\label{eq: eff HN}
\eeq
Here, $J_{\rm eff}$ is the effective tunneling strength for a pair,
\beq
J_{\rm eff} = \frac{J^2}{U_0}\,,
\eeq
and $H^t$ is the tunneling contribution,
\beq
H^t = - \frac{J^{\rm eff}}{2} \sum_{<i,j>} (T_{i,j} + T_{j,i})
\eeq
with $T_{i,j}$ the 
operators that correspond to hopping of a pair from site $i$ to $j$. The interactions between nearest-neighbours are contained in a contribution $H^d$ that will be diagonal in our basis and is of the form:
\beq
H^d = J^{\rm eff} \sum_{<i,j>} N_i N_j \,. 
\eeq
In this expression, $N_j$ is the number of pairs in site $j$, with double occupations being forbidden
because of the fermionic character of the constituents. 

Our strongly bound fermion pairs are thus the same as hard-core bosons with a specific relation between 
tunneling strength and nearest-neighbour repulsion. The hopping term tends to delocalize the cobosons, whereas the repulsive interaction together with the hard-core character 
compete with the hopping. Therefore one expects that the ground state will have 
delocalized pairs but which are unlikely to be found next to each other. We note that previous work \cite{lasmar_2019} has considered the inclusion of longer-ranged attractive interactions that can lead to the formation of larger aggregates. Such generalizations imply a richer variety of behaviours and are beyond the scope of the present study.

From Eq.~(\ref{eq: eff HN}) we can write down the matrix form of the Hamiltonian for any particular subspace with fixed $N$. For brevity, in the following we ignore the overall energy $-N (U_0+ J^{\rm eff})$. For $N=1$ the basis of our restricted Hilbert space is composed by states of the form $\ket{k}$ where $k$ labels the location of the bound pair in the graph, and the Hamiltonian contains only tunneling:
\begin{equation}
 H_{k,l}^t = -\frac{J_{\rm eff}}{2} A_{kl} \,.
\end{equation}
Here $A_{kl}$ is the corresponding element of the adjacency matrix, which is 1 when sites $k$ and $l$ are neighbours and vanishes otherwise. 

For $N=2$ the basis of the Hilbert space is composed by the states $\ket{k,l}$ where $k,l$ label the locations of each of the pairs in the graph, and using Pauli exclusion principle we can take $k<l$. In this basis, the diagonal part is:
\begin{equation}
 H_{kl,kl}^d = J_{\rm eff} A_{kl}
\end{equation}
while the tunneling part of the Hamiltonian is of the form:
\begin{equation}
 H_{kl,mn}^t = -\frac{J_{\rm eff}}{2} (\delta_{ln} A_{km} + \delta_{lm} A_{kn} + \delta_{kn} A_{lm} + \delta_{km} A_{ln} ) \,. 
\end{equation}

For $N=3$ the basis is of the form $\ket{k,l,m}$ with $k<l<m$. The diagonal part of the Hamiltonian is:
\begin{equation}
 H_{klm,klm}^d = J_{\rm eff} ( A_{kl} + A_{km} + A_{lm} )
\end{equation}
whereas the tunneling term can be written as:
\begin{equation}
 H_{ijk,lmn}^t = -\frac{J_{\rm eff}}{2} ( A_{il} \delta_{jm} \delta_{kn} + \text{all permutations} ) 
\end{equation}
where by ``all permutations'' we mean all permutations of indices $i,j,k$ and $l,m,n$ separately.

\section{Coboson ansatz for the ground state}\label{sec:cobosons}

In this work we focus on  identical composite bosons, each made of two distinguishable 
fermions. This section provides a brief overview of the coboson ansatz for the 
ground state of $N$ such pairs. For a more complete 
introduction to the coboson formalism, we refer the reader to \cite{combescot_2008}.
For a given Hamiltonian corresponding to a single pair, the ground state $|\psi\rangle$
defines the coboson creation operator $B^\dagger$, namely the operator which acts on 
the vacuum state $\ket{v}$ creating a single pair in the ground state, $|\psi\rangle=B^\dagger|v\rangle$. 

One can write a normalized state of $N$ 
composite bosons obtained after acting $N$ times with the coboson creation 
operator in the form \cite{law_2005,combescot_2008}
\begin{equation}
\ket{N} = \frac{\left( B^\dagger \right)^N}{\sqrt{ N! \chi_{N} } } \ket{v}.
\label{eq:ansatz}
\end{equation}
Here $\chi_{N}$ is a normalization factor which accounts for the Pauli exclusion principle \cite{law_2005, 
combescot_2003}.

For composite bosons made of two fermions the normalization 
factor depends on the Schmidt coefficients $\lambda_j$ of $\ket{\psi}$ and takes the form \cite{law_2005, combescot_2003, chudzicki_2010},
\begin{equation}
\label{chiN}
\chi_{N} = N! \sum_{p_N > p_{N-1} > \text{...}> p_1} \lambda_{p_1} 
\lambda_{p_2} \text{...}\lambda_{p_N}\,.
\end{equation}
For the case $N=2$, the normalization coefficient is equal to $\chi_2 = 1-P$, 
with $P$ the purity of the reduced density matrix of one of the constituent 
particles of a pair in the ground state $|\psi\rangle$. Thus, a high entanglement in state $|\psi\rangle$ is a key aspect behind effectively bosonic features. In general, the behavior of pairs as approximate elementary bosons can be related to the normalization 
coefficients, and bosonic behavior is recovered when 
$\chi_N/\chi_{N-1}\simeq 1$ \cite{law_2005, chudzicki_2010, combescot_2011b,tichy_2012b}.

In particular, for the model we consider the purity $P$ is determined by the distribution of occupation probabilities $p_1(j)$ of the various sites $j$ in the single-pair ground state, according to:
\begin{equation}
    P = \sum_{j=1}^M p_1^2(j)\,.
\end{equation}
For a fixed number of sites, the maximum entanglement or minimal purity corresponds to a uniform probability distribution.

The idea that the state $\ket{N}$ of Eq.~\eqref{eq:ansatz} provides a good approximation of the ground state of a system of $N$ cobosons is an important element of coboson theory, and we 
refer to this in the following as the coboson ansatz. This ansatz provides an enormous simplification of the description of the many-body ground state, in the cases when it is applicable. Nonetheless, we note that the coboson formalism is a powerful machinery also when this simplification is not possible, as illustrated by several examples in the literature, see for instance \cite{combescot_2015, Cuestas_Cormick_2022, Jimenez_2023}.

A useful property of state $\ket{N}$ is that for dilute systems one can derive simple expressions for the expectation values of many observables of interest in terms of an expansion in the density of particles. Following the steps in \cite{combescot_2008}, one obtains approximations of the form:
\beq
\langle O \rangle_N \simeq N \langle O \rangle_1 + \frac{N(N-1)}{2} (\langle O \rangle_2-2\langle O \rangle_1)
\label{eq:expansion}
\eeq
up to corrections of third order in particle density. Here, $\langle O \rangle_N$ denotes the mean value of operator $O$ evaluated over state $\ket{N}$. The derivation in \cite{combescot_2008} takes $O$ to be the system Hamiltonian, but the same steps can be carried out for any observable which can be written as a one-body operator either in terms of cobosons or in terms of elementary fermions.

Problems with full translational invariance, like the ring and torus studied in \cite{Cespedes_2019}, lead to very simple forms for the state $\ket{N}$. The ground state of a single pair in such a lattice is the same as in the fully connected graph, i.e.:
\beq
\label{eq:phi_symm}
\ket{\psi} = \frac{1}{\sqrt{M}} \sum_{j=1}^M a^\dagger_j b^\dagger_j \ket{v} \,.
\eeq
The structure of this state greatly simplifies many calculations, because 
all Schmidt coefficients of this state are equal to $1/M$. For states of the form of Eq.~(\ref{eq:phi_symm}) the coboson ansatz has full symmetry between all sites for arbitrary $N$.

In general, the coboson ansatz is expected to be a satisfactory approximation when the system is very dilute, interactions are sufficiently short-ranged, and the single-pair ground state $\ket{\psi}$ exhibits high entanglement between the pair constituents. Previous work, however, has shown that the validity of the ansatz may also depend on the dimensionality \cite{Cespedes_2019}. This is to be expected since the coboson ansatz does not properly capture the strong spatial correlations among pairs that build up for one-dimensional or quasi one-dimensional systems.

\section{Graph definitions and relevant properties}
\label{sec:graphs}

In the light of the previous results, in this work we consider more general graphs with both integer and non-integer dimensions, to better study the interplay between dimensionality and applicability of the coboson ansatz. Before presenting our findings, in the following we introduce the fractal graphs that we will consider, as well as some concepts of graph theory that will be important for our analysis.

In the next Section we numerically examine the ground state of two and three fermion pairs in different graphs with dimensions between 1 and 2, considering the Hausdorff dimension for the case of fractals \cite{Falconer_2003}. In contrast with regular lattices, which have integer dimensions, the Hausdorff dimension can take non-integer values to describe the geometry of fractal graphs. It is related to the box-counting dimension, associated with the growth of the number $N(r)$ of balls of radius at most $r$ required to cover a graph region completely, in the limit when $r$ goes to zero. 

We note that fractal dimension is only well defined in the limit of infinitely many points in the graph, whereas our systems are always of finite size. Our interest, however, is precisely the behavior of the ground state as the infinite-size limit is approached and the system becomes more dilute. We regard fractal dimension as a good measure for our purposes since the increase in system size in our study is achieved by considering successive levels of the iterative construction of the fractals.

\begin{figure}[t]
 \includegraphics[width=\columnwidth]{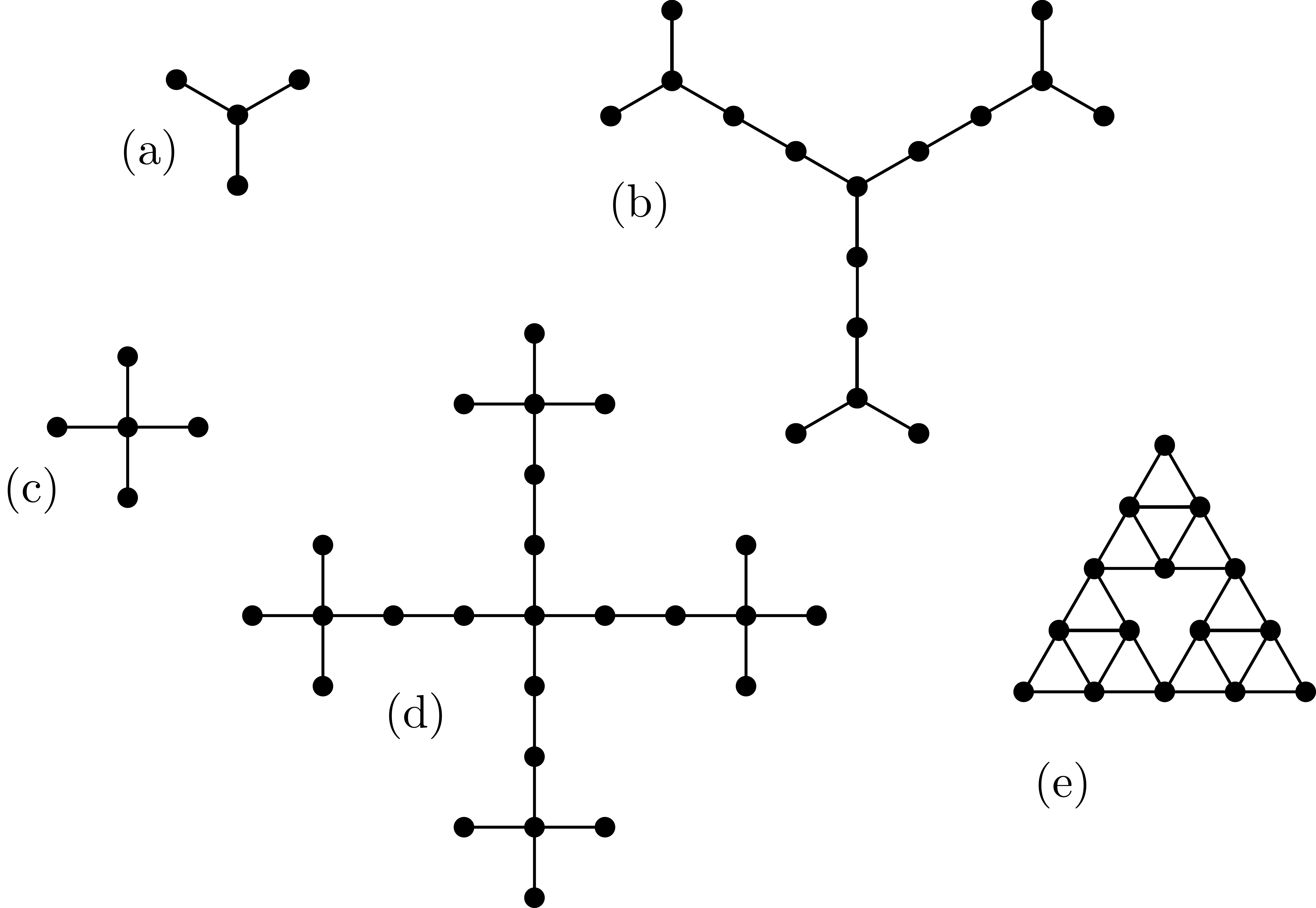}
  \caption{Small-size versions of the fractal graphs we consider. a) and b) First and second levels of the Vicsek fractal with $\nu=3$ respectively. c) and d) First and second levels of the Vicsek fractal with $\nu=4$. e) Sierpinski gasket with 15 sites. \label{fig:graphs}}
\end{figure}

We take the following instances of graphs: 1) one-dimensional lattices given by simple chains (which we label 1D); 2) two-dimensional square lattices (2D); 3) fractal graphs given by the Sierpinski gasket, also called Sierpinski triangle, with dimension $\log_2(3)\simeq1.58$ and which we label S; 4) Vicsek fractals, with dimensions $d = \log(\nu + 1)/ \log(3) \simeq 1.26$ for $\nu=3$ and $\simeq 1.47$ for $\nu =4$  \cite{Vicsek_1983, Blumen_2004}, labelled V3 and V4 respectively. Both the Sierpinski and the Vicsek fractals are built iteratively, so that the actual fractal structure corresponds to the limit of infinitely many iterations. The step-wise construction of these fractals is illustrated in Fig.~\ref{fig:graphs}. In the case of the Vicsek fractals, the parameter $\nu$ is associated with the number of copies of the graph that are appended to it to build the next iteration, and the first level is a star graph with $\nu$+1 points. 

\begin{figure}[t]
  \includegraphics[width=\columnwidth]{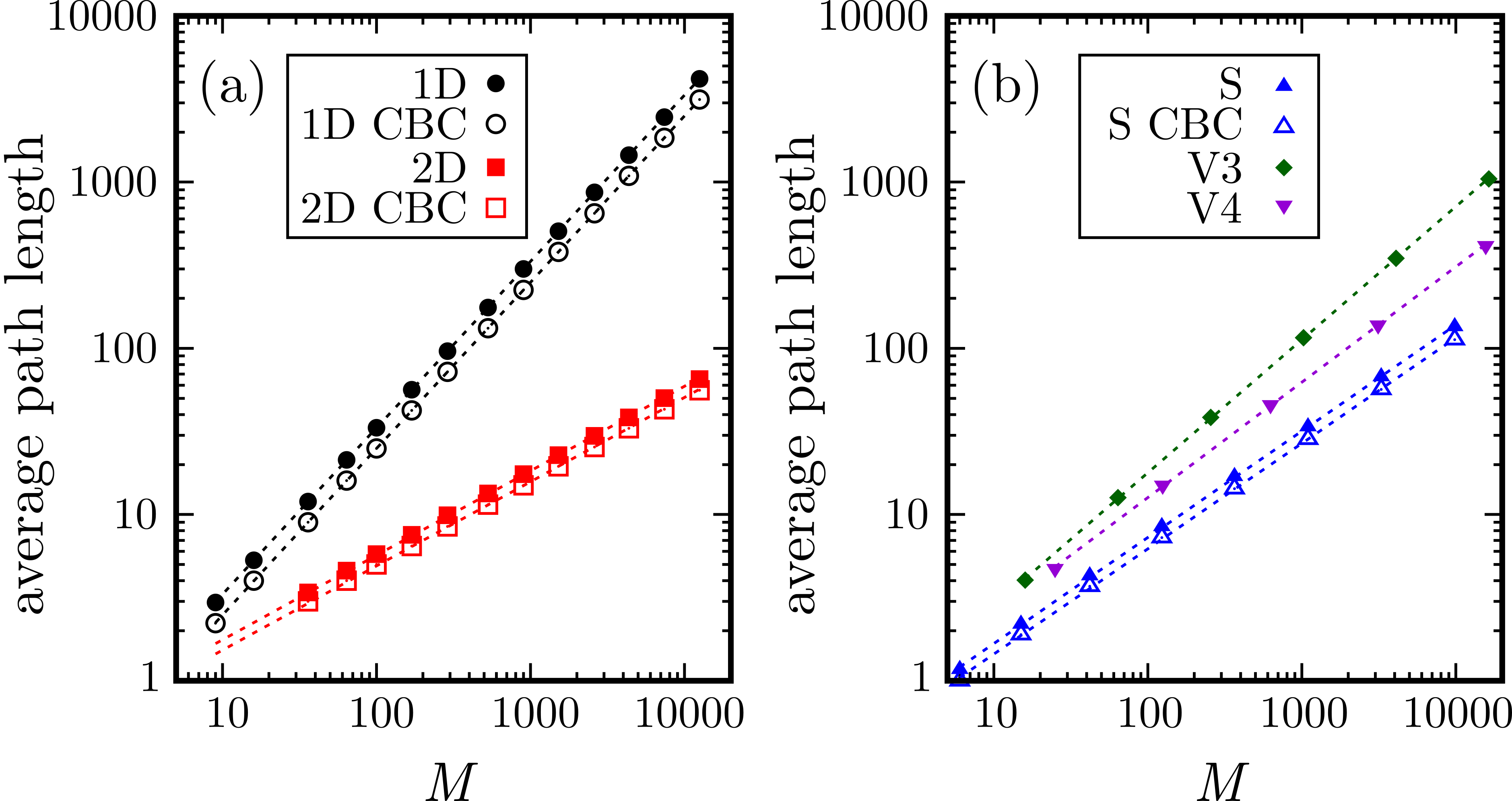}
  \caption{Growth of the average path length between sites (in logarithmic scale) as a function of the total number $M$ of sites in the graph, for: a) 1D chains, 2D square lattices, and b) Sierpinski gaskets and Vicsek fractals with $\nu=3,4$. For regular lattices and Sierpinski gaskets we consider both open and closed boundary conditions, the latter labelled by ``CBC''. In each case the inverse of the power of the growth of the average path length, $\alpha$, obtained from the numerical fit and represented with dotted line, coincides with the graph dimension.\label{fig:MDG}}
\end{figure}

The dimension of the graphs we consider is related to the growth of the average path length between graph nodes as the size of the graph is increased. Here, the path length is taken to be the number of edges along the shortest path between nodes; this quantity is then averaged over all pairs of nodes. In Fig.~\ref{fig:MDG} we show the growth of the average path length for the system sizes we explore. For all graphs except the Vicsek fractals we consider both open and closed boundary conditions; for the Sierpinski gaskets the modified boundary conditions are obtained adding extra edges linking the outer vertices of the triangle. As can be seen in the plots, the numerical fits of the growth of the average path length indicate a power that coincides with the inverse of the graph dimension and that is independent of the boundary conditions.

Apart from the dimensionality, the quality of the coboson ansatz may also be affected by other graph properties. Two concepts from graph theory that will be relevant in the following are those of circuit rank and betweenness. The circuit rank of a connected graph is equal to the minimum number of edges that must be removed to turn the graph into a tree \cite{Berge_2001}, and so in comparison with the total number of edges, it gives an idea of how close the graph is to being a tree. For instance, the circuit rank is equal to zero for a tree, whereas it grows quadratically with the number of nodes for a fully connected graph and linearly for a regular lattice of dimension larger than 1.

The betweenness centrality or betweenness $g$ of each node in a graph is a measure of its centrality in the graph \cite{newman2018networks}. It is related to the fraction of shortest paths among graph nodes that pass through the particular vertex considered.
Among the graph families we study, the distribution of betweenness of the different nodes varies significantly. For the graphs with closed boundary conditions, all sites have the same value of $g$; nontrivially, this is also valid for the Sierpinski gasket. Among the graphs with open boundary conditions, regular lattices have nodes with larger betweenness at the center, but the variation is smooth. In contrast, for Vicsek fractals the central nodes have values of $g$ that are much larger than those of the rest. 

Actually, betweenness centrality is related to another aspect that is relevant to our study, namely the purity $P$ of the reduced density matrix of each fermion in the single-pair ground state, which depends on the kind of graph considered. As explained in the previous section, bosonic behaviour of the coboson creation operator is directly associated with $P$. To take this into account, we characterized the effective size $S$ of our system as $S=1/P$ for each given lattice. The reason to consider $S$ as an effective size is that a value of $S$ smaller than $M$ implies that a pair in the ground state is not spread over all sites equally, and we wish to correct for this effect when we study the validity of the coboson ansatz. 

\begin{figure}[h]
    \centering
    \includegraphics[width=\columnwidth]{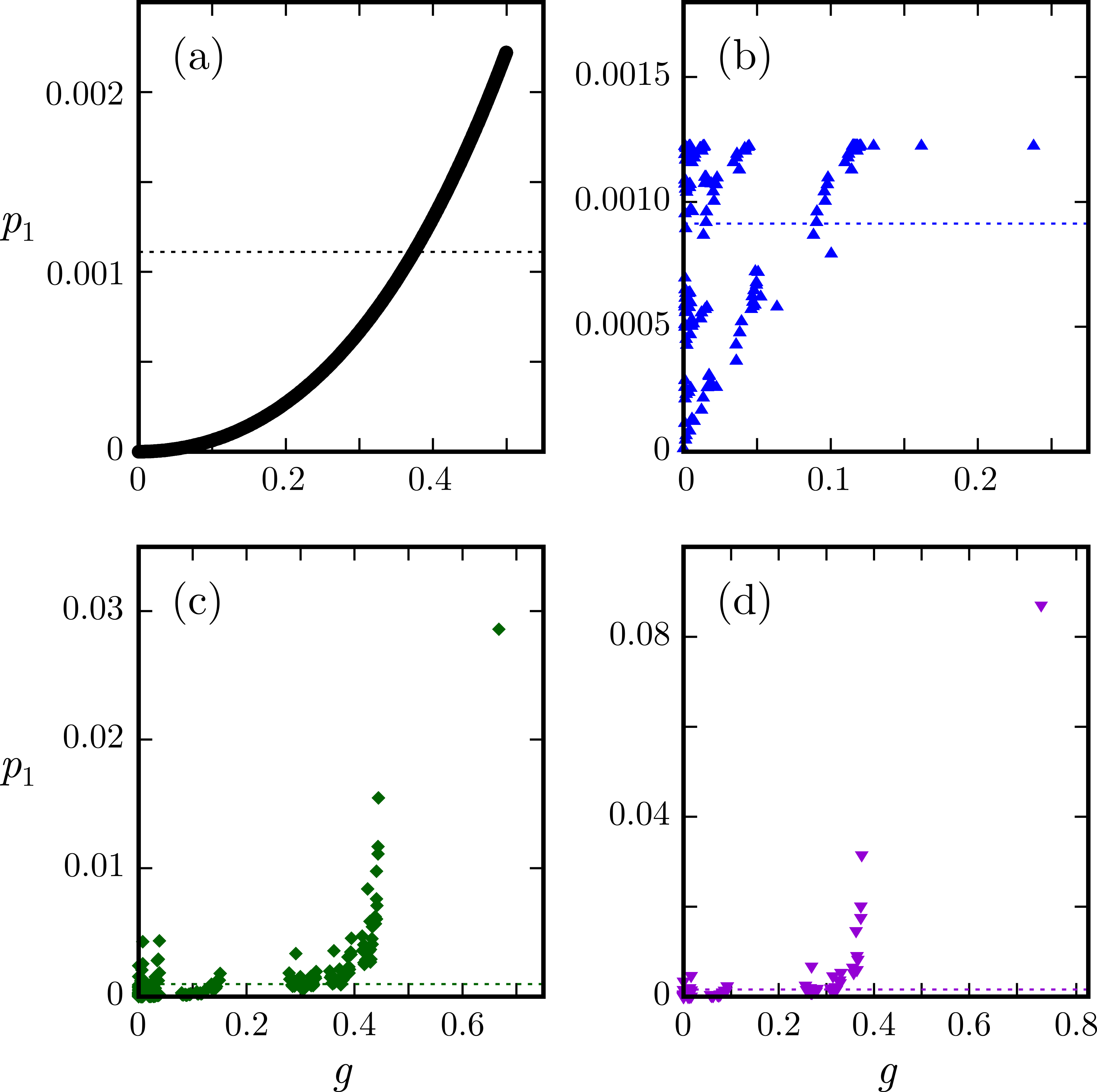}
    \caption{Occupation probability $p_1$ of each site in the single-pair ground state vs. centrality of the node, as quantified by the betweenness centrality $g$, for graphs with open boundary conditions and numbers of sites $M\sim1000$. The subplots correspond to: a) 1D chain with $M=900$, b) Sierpinski gasket with $M=1095$, c) Vicsek fractal with $\nu=3$ and $M=1024$, and d) Vicsek fractal with $\nu=4$ and $M=900$. Dashed horizontal lines represent values of uniform probability given by $1/M$.
    \label{fig:betweenness}}
\end{figure}

It turns out that, although there is no one-to-one correspondence, the single-pair ground state tends to localize at the nodes with larger betweenness, as illustrated in Fig. \ref{fig:betweenness}. We note that the 2D case is not shown in Fig.~\ref{fig:betweenness} but looks similar to the 1D chain. The very uneven occupation probability found for the Vicsek fractals can be better appreciated observing the dashed horizontal lines that indicate the average over all sites. 

\begin{figure}[h]
  \includegraphics[width=\columnwidth]{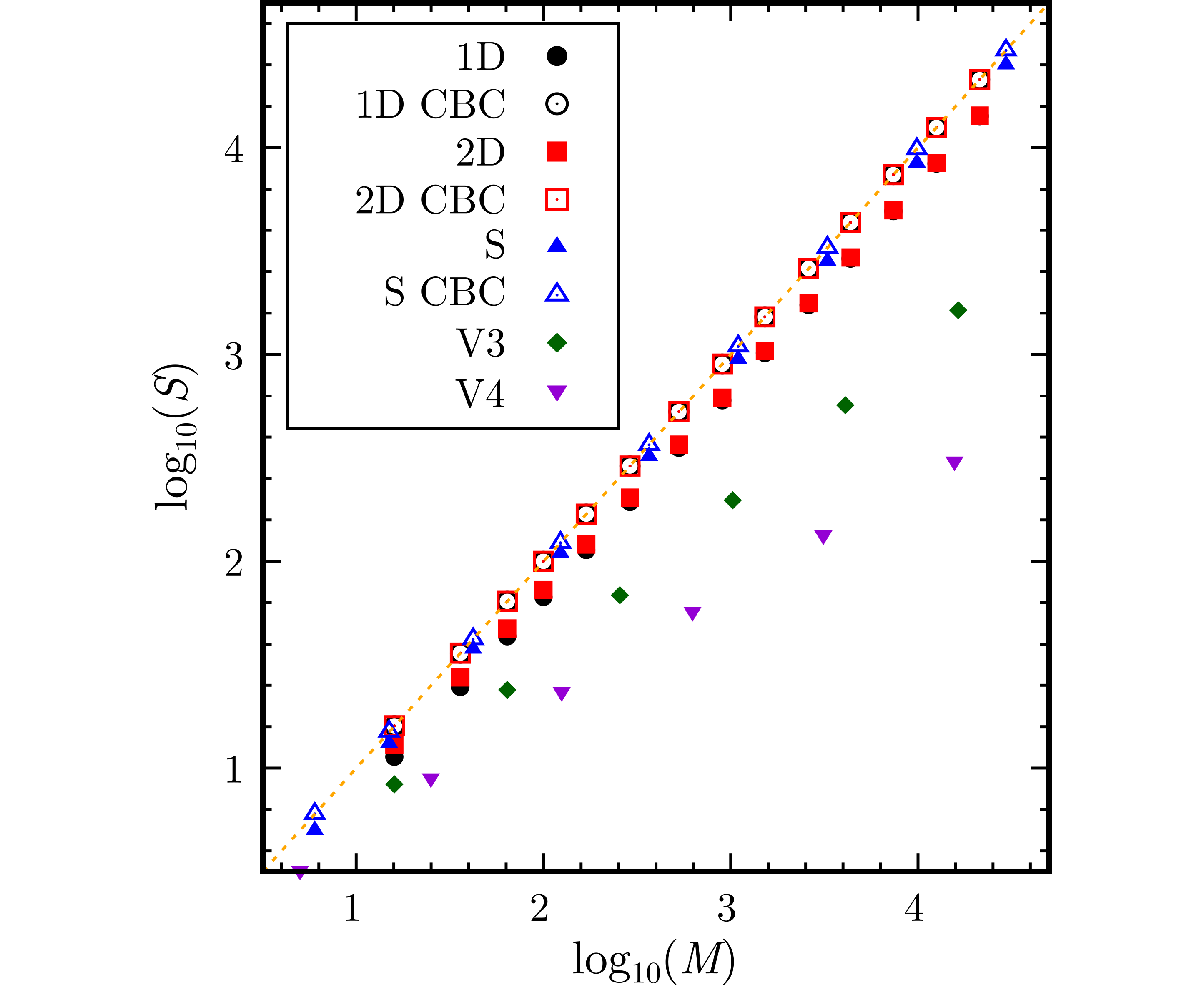}
  \caption{Effective size quantified by log$_{10}$($S$) as a function of log$_{10}$($M$) for the graphs studied, with $S=1/P$ the effective system size and $M$ the total number of sites. The dashed line shows the identity $M=S$ that is obtained for uniformly distributed single-pair ground states. The lowest values of $S(M)$ are obtained for Vicsek fractals. \label{fig:MS}}
\end{figure}

Thus, graphs with very unbalanced betweenness values also have larger differences between the number of sites $M$ and the effective size $S$. The relation between $M$ and $S$ for various graphs is displayed in Fig.~\ref{fig:MS}. Indeed, the lowest values of $S$ for a given $M$ are found for Vicsek fractals. The limit value $S=M$ is shown with a dashed line and is found for the one-dimensional ring and the torus (i.e. the square lattice with periodic boundary conditions). Closed boundary conditions always lead to larger values of $S$ for fixed $M$, thus implying larger effective sizes without increasing the numerical cost. However, for large systems the relative difference in effective size becomes rather small, as boundary effects become less important.

\section{Fidelity between the ground state and the coboson ansatz for various geometries}

\subsection{Preliminary considerations}

The goal of this study is to test the validity of the ansatz in different graphs with dimensions between 1 and 2. For this purpose, we first find the ground state of a single composite particle in the lattice, of the form:
\begin{equation}
 \ket{\phi_{GS}^{(1)}} = \sum_{j=1}^M c_j \ket{j} \,.
 \label{eq:state1pair}
\end{equation}
where $\ket{j}$ labels the location of the pair on the lattice. Because the two fermions forming a pair are always at the same site, the Schmidt decomposition is trivial, with Schmidt coefficients given by $\lambda_j=c_j^2$, since for our model $c_j\in\mathbb{R}$. 

From state (\ref{eq:state1pair}), we can build the coboson ansatz for an arbitrary number of pairs. Our aim is to assess the quality of the coboson ansatz for the ground state in different graphs, and we do so by computing the fidelity between the ground state $\ket{\phi_{GS}^{(N)}}$ obtained numerically and the coboson ansatz $\ket{N}$ according to:
\beq
\mathcal{F}_N = \left|\braket{N}{\phi_{GS}^{(N)}}\right|^2 \,.
\eeq
If the coboson ansatz is applicable, the fidelity $\mathcal{F}_N$ is expected to approach 1 in the infinitely dilute limit, i.e. when the number of sites $M$ tends to infinity. We note that $\mathcal{F}_N$ depends on both $N$ and $M$ and also on the kind of graph considered; this dependence will be left implicit to keep the notation simple.

One could consider, instead of calculating the fidelity, a comparison of the ground-state energies obtained from numerical calculation and from the coboson ansatz. A clear disagreement between these values would already imply a failure of the ansatz. However, as has been shown in \cite{Cespedes_2019}, it is possible that the energies agree to a good approximation while at the same time other properties such as spatial correlations are not correctly reproduced.

In the following we will focus on the cases $N=2,3$. For two composite particles we get:
\begin{equation}
 \ket{N=2} = \sqrt{\frac{2}{\chi_2}} \sum_{j<k} c_j c_k \ket{j,k} \,.
\end{equation}
If the ground state for two pairs has coefficients given by
\begin{equation}
 \ket{\phi_{GS}^{(2)}} = \sum_{j<k} \alpha_{jk} \ket{j,k}
\end{equation}
then the fidelity takes the form:
\beq
\mathcal{F} = \frac{2}{\chi_2} \left(\sum_{j<k} \alpha_{jk} c_j c_k\right)^2
\eeq
where we are using that all coefficients are real.

Along the same lines, for the case of three pairs, $N=3$, the expression for the fidelity takes the form:
\beq
\mathcal{F}_3 = \frac{6}{\chi_3} \left(\sum_{j<k<l} \alpha_{jkl} \, c_j c_k c_l \right)^2 
\eeq
with $\alpha_{jkl}$ the coefficients of the ground state for $N=3$ in the basis $\ket{j,k,l}$.

A few cases can be readily analyzed: the fully connected graph, due to its symmetry, has the coboson ansatz as exact ground state, and thus $\mathcal{F}_N=1$ for any number of sites as long as it can acommodate $N$ pairs. Another example that is solvable because of its high symmetry is the star graph. 
We note, however, that the creation operator of the single-pair ground state on the star cannot exhibit proper bosonic behaviour, because $\chi_N/\chi_{N-1}$ does not tend to 1 as the number of sites goes to infinity.

The 1D chain with closed boundary conditions has been studied in \cite{Cespedes_2019}. The single-pair ground state has the fully symmetric form of Eq.~\eqref{eq:phi_symm}, and the fidelity for $N=2$ with $M\to\infty$ is equal to $8/\pi^2 \simeq 0.81$, as has been shown using the mapping to the Heisenberg model. A 1D chain with open boundaries has a non-uniform single pair ground state, and the asymptotic fidelity for $N=2$ is $2^{17}/(\pi^4 45^2)\simeq 0.66$ as can be seen using fermionization~\cite{Girardeau_1960}. Thus, not surprisingly, 1D systems do not exhibit condensation in the dilute limit and the coboson ansatz does not approach the ideal fidelity of 1.

Previous work \cite{Cespedes_2019} has also numerically studied 2D square lattices with periodic boundary conditions, observing a fidelity which increases monotonically with system size and suggesting an asymptotic value of $\mathcal{F}_2=1$. On the contrary, the fidelity for ladder systems, i.e. $n\times m$ lattices with $n\to\infty$ and $m$ fixed, has a behaviour that resembles that of 1D chains. It seems then that the effective dimension of the lattice for our purposes is best captured by the power of the growth of the average path length between sites: ladder models, indeed, have average path lengths that asymptotically grow linearly in the total number of sites, just as 1D systems. It is natural to conjecture that an average path length growth with power 1 is associated with the failure of the coboson ansatz.

\subsection{Numerical results}
\label{sec:results}

In the following we analyze the behavior of the fidelity between the coboson ansatz and the numerical ground state for increasing sizes of graphs with different geometries. We note that, as observed in \cite{Cespedes_2019}, the convergence of the fidelity with increasing system size is rather slow. The study of large systems in \cite{Cespedes_2019} was largely simplified by exploiting translational invariance, which is no longer possible in the present work. 

\begin{figure}[h]
  \includegraphics[width=\columnwidth]{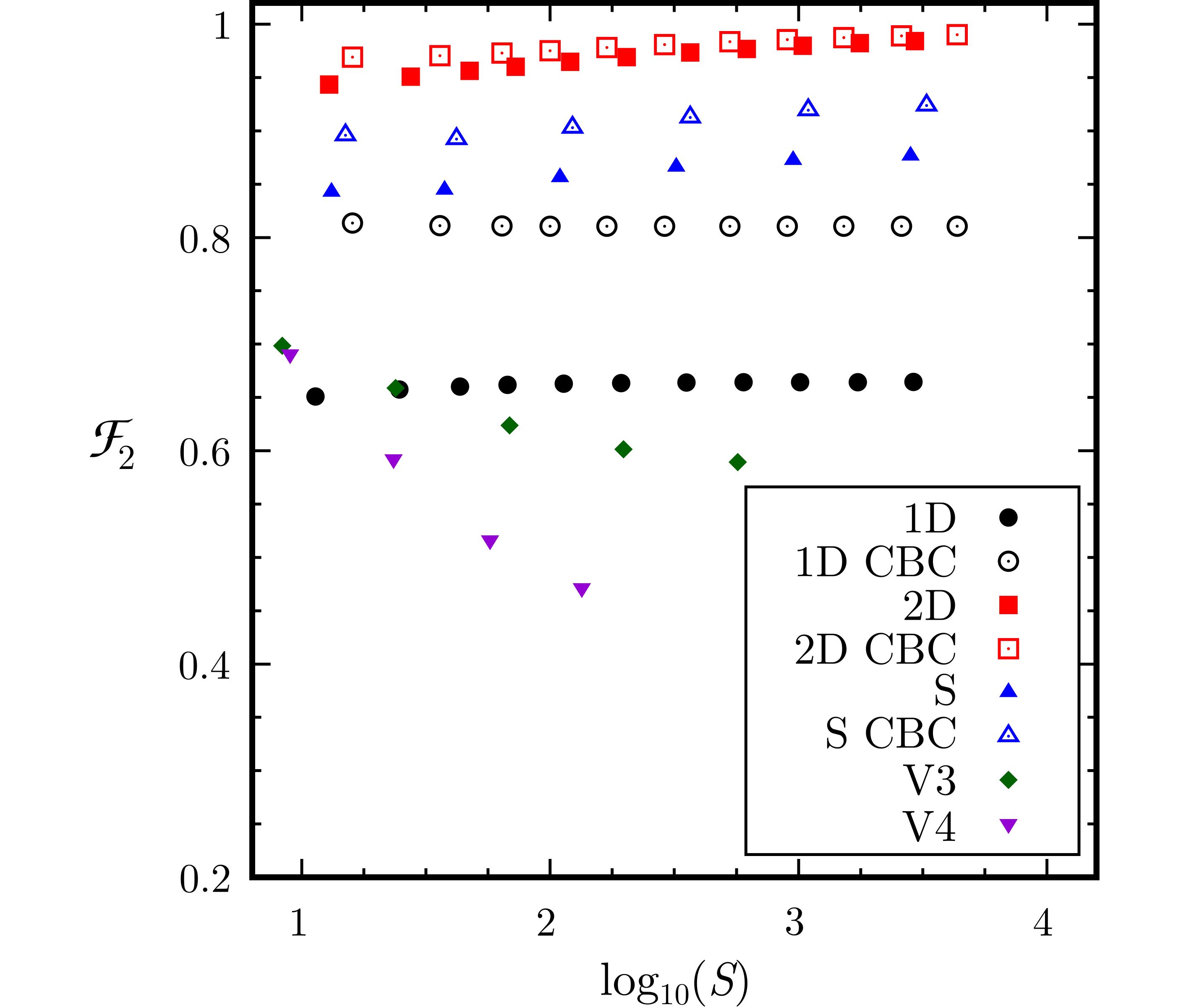}
  \caption{Fidelity between the coboson ansatz and the true ground state for two pairs, quantified by $\mathcal{F}_2$ as a function of $\log_{10}(S)$ for different graphs.
  \label{fig:SF2}}
\end{figure}

In Fig.~\ref{fig:SF2} we show the fidelity for two pairs, $\mathcal{F}_2$, as a function of $S$ for the different networks considered. We note that apart from the cases plotted and already described, we performed calculations with open square, triangular and hexagonal 2D lattices, leading to very similar results as the open square lattice. Calculations with the Hanoi graph, or dual Sierpinski gasket, also lead to very similar results as the Sierpinski gasket. Just as the various geometries for two-dimensional lattices, the Hanoi graph and the Sierpinski gasket have the same dimension but different number of neighbours. From these comparisons, we conclude that the graph degree does not play a relevant role for our purposes.

Several features of Fig.~\ref{fig:SF2} are interesting: in general, the plots always show a higher fidelity for closed boundary conditions (empty symbols), even after correcting for the effective size using $S$ instead of $M$. Leaving aside the case of the Vicsek fractals, a higher dimension is associated with a higher fidelity, and graphs with dimension higher than 1 seem to have fidelities that improve as the graph size is increased. Of course, this does not imply that the fidelity approaches the ideal value of 1 as the number of sites approaches infinity; it may well be that convergence is observed at much larger values than the ones numerically accessible.

The case of the Vicsek fractals is strikingly different from the other graphs studied; indeed, in Vicsek fractals the coboson ansatz actually gets worse as the system size is increased, and the corresponding fidelities lie clearly below the case of 1D chains. We associate this behaviour with the tree-like character of the Vicsek fractals (see Fig.~\ref{fig:graphs}). We thus conjecture that the validity of the coboson ansatz for dilute systems in general graphs is to be expected when not only the average path length grows with a power lower than 1, but also the graph should have an unbounded circuit rank. 

\begin{figure}[h] \includegraphics[width=\columnwidth]{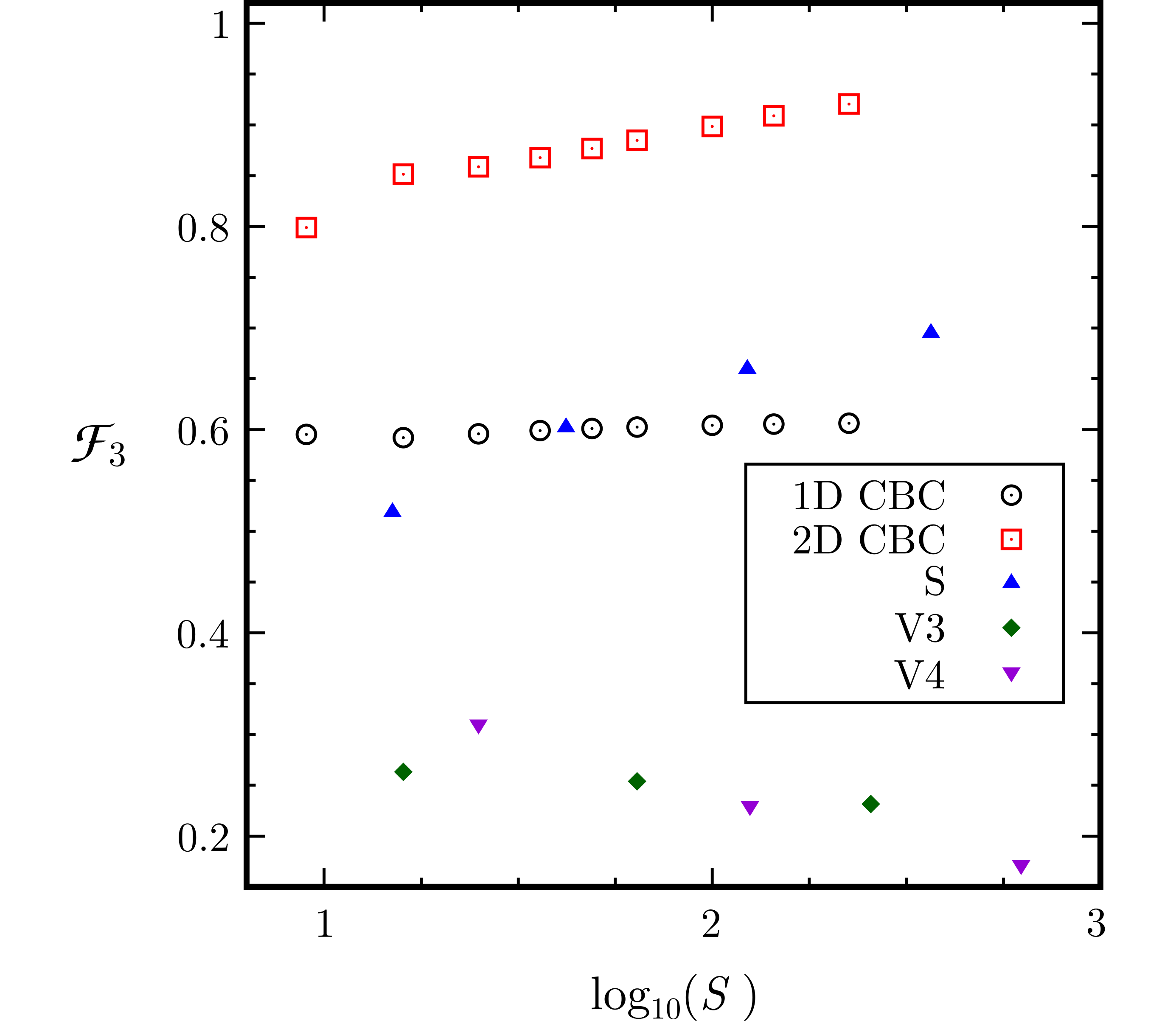}
  \caption{Fidelity between the coboson ansatz and the true ground state for three pairs, quantified by $\mathcal{F}_3$ as a function of log$_{10}$($S$) for different graphs.
  \label{fig:SF3}}
\end{figure}

As explained in Sec.~\ref{sec:cobosons}, a failure of the coboson ansatz for $N=2$ generally leads to wrong predictions for higher numbers of particles. However, as a further check, we also study the fidelity for increasing graph size for systems with $N=3$. In this case, we restrict to smaller graphs to keep the full Hilbert space tractable. The results are shown in Fig.~\ref{fig:SF3}, and indeed display a similar behaviour as the previous figure: the fidelity increases with graph size for the 2D lattice and the Sierpinski gasket, whereas it exhibits convergence to a value clearly below 1 for the 1D chain and it decreases for the Vicsek fractals.

There is actually an intuitive explanation why tree-like graphs can be expected to behave very differently from other graphs. Indeed, the presence of a first pair on one site blocks that site for a second pair; thus one can think of that second pair as moving in a graph which is modified by the removal of the blocked site. In the case of 2D graphs, it is clear that this effect is minor; on the contrary, for tree-like graphs removing a single site generally splits the graph in two or more isolated parts. 

Along this line, one can then think that the key feature determining whether the ground state can have condensate-like properties is given by the vertex-connectivity, i.e. the minimum number of nodes that have to be removed to make the graph disconnected. This, however, does not seem to be the case, since Sierpinski triangles have low connectivity and their behaviour is more similar to 2D graphs than to Vicsek fractals. There is also a reason why the connectivity is not necessarily relevant: it is not only important that removing a node will modify the properties of the graph; it also matters how likely it is to find a pair at a particular given site. 

This point can be better examined considering once more the probability distribution $p_1$ of occupation of the various nodes for the single-pair ground state as displayed in Fig.~\ref{fig:betweenness}. For the case of the Vicsek fractals, we notice a few sites have an occupation probability that is way larger than the average. These sites are precisely those whose removal has the largest impact in the properties of the remaining graph, in particular the central node. This is not the case in the Sierpinski gasket, or in 1D and 2D lattices, for which many nodes have comparably high occupation. We conjecture that this strong relation between centrality and occupation probability makes tree-like graphs particularly unfavorable for condensation.

\section{Conclusions}
\label{sec:conclusions}

We have studied a system of a few ($N=2,3$) strongly bound fermion pairs in graphs of dimension between 1 and 2. In order to characterize the behaviour of the composite particles at zero temperature, we have examined the fidelity between the ground state found numerically and the so-called ``coboson ansatz''. The latter is a condensate-like state but taking into account the fermionic character of the constituents of each pair. Although we restricted to systems with 2 or 3 pairs, the structure of coboson theory is such that at low densities many expectation values are dominated by the results for few pairs, and so our conclusions are relevant also for the many-body scenario \cite{combescot_2008}.

Our results indicate a poor performance of the coboson ansatz for 1D chains and Vicsek fractals. From our findings we conjecture that condensation of fermion pairs in the very dilute limit can be expected when: i) the average path length grows with a power strictly lower than 1, and ii) the graph has unbounded circuit rank and is thus ``very different from a tree''. These conditions should be considered together with the already known criteria associated with the entanglement between the constituents of a single pair in its ground state and the short-ranged interactions.

Among the graphs fulfilling conditions i) and ii) above, we find a strong correlation between the dimension and the quality of the coboson ansatz: the larger the dimension, the better the description provided by the ansatz. Boundary conditions also play an important role: lattices with closed boundary conditions consistently display larger fidelities than their open-boundary counterparts. On the other hand, the comparison of different kinds of 2D lattices, and of the Sierpinski gasket with its dual, suggests that the number of neighbours does not affect the results significantly. 

We note that the same conclusions apply to standard hard-core bosons, since we observed little difference when the nearest-neighbour interactions resulting from Pauli exclusion were removed. Given the slow convergence of the fidelity with system size, it would be desirable to extend our analysis to larger graphs. This will require more sophisticated numerical techniques; direct diagonalization is too costly due to the quadratic scaling of the Hilbert space and the lack of translational invariance. We hope that the present work sparks further interest in the problem of condensation of composite particles in arbitrary geometries.

\section{Acknowledgments}

The authors acknowledge funding from grants PICT 2017-2583, PICT 2018-02331 and PICT 2020-SERIEA-00959 from ANPCyT (Argentina).


\begin{thebibliography}{45}
\expandafter\ifx\csname natexlab\endcsname\relax\def\natexlab#1{#1}\fi
\expandafter\ifx\csname bibnamefont\endcsname\relax
  \def\bibnamefont#1{#1}\fi
\expandafter\ifx\csname bibfnamefont\endcsname\relax
  \def\bibfnamefont#1{#1}\fi
\expandafter\ifx\csname citenamefont\endcsname\relax
  \def\citenamefont#1{#1}\fi
\expandafter\ifx\csname url\endcsname\relax
  \def\url#1{\texttt{#1}}\fi
\expandafter\ifx\csname urlprefix\endcsname\relax\def\urlprefix{URL }\fi
\providecommand{\bibinfo}[2]{#2}
\providecommand{\eprint}[2][]{\url{#2}}

\bibitem[{\citenamefont{Endres et~al.}(2016)\citenamefont{Endres, Bernien,
  Keesling, Levine, Anschuetz, Krajenbrink, Senko, Vuletic, Greiner, and
  Lukin}}]{Endres_2016}
\bibinfo{author}{\bibfnamefont{M.}~\bibnamefont{Endres}},
  \bibinfo{author}{\bibfnamefont{H.}~\bibnamefont{Bernien}},
  \bibinfo{author}{\bibfnamefont{A.}~\bibnamefont{Keesling}},
  \bibinfo{author}{\bibfnamefont{H.}~\bibnamefont{Levine}},
  \bibinfo{author}{\bibfnamefont{E.~R.} \bibnamefont{Anschuetz}},
  \bibinfo{author}{\bibfnamefont{A.}~\bibnamefont{Krajenbrink}},
  \bibinfo{author}{\bibfnamefont{C.}~\bibnamefont{Senko}},
  \bibinfo{author}{\bibfnamefont{V.}~\bibnamefont{Vuletic}},
  \bibinfo{author}{\bibfnamefont{M.}~\bibnamefont{Greiner}}, \bibnamefont{and}
  \bibinfo{author}{\bibfnamefont{M.~D.} \bibnamefont{Lukin}},
  \bibinfo{journal}{Science} \textbf{\bibinfo{volume}{354}},
  \bibinfo{pages}{1024} (\bibinfo{year}{2016}).

\bibitem[{\citenamefont{Schymik et~al.}(2020)\citenamefont{Schymik, Lienhard,
  Barredo, Scholl, Williams, Browaeys, and Lahaye}}]{Schymik_2020}
\bibinfo{author}{\bibfnamefont{K.-N.} \bibnamefont{Schymik}},
  \bibinfo{author}{\bibfnamefont{V.}~\bibnamefont{Lienhard}},
  \bibinfo{author}{\bibfnamefont{D.}~\bibnamefont{Barredo}},
  \bibinfo{author}{\bibfnamefont{P.}~\bibnamefont{Scholl}},
  \bibinfo{author}{\bibfnamefont{H.}~\bibnamefont{Williams}},
  \bibinfo{author}{\bibfnamefont{A.}~\bibnamefont{Browaeys}}, \bibnamefont{and}
  \bibinfo{author}{\bibfnamefont{T.}~\bibnamefont{Lahaye}},
  \bibinfo{journal}{Phys. Rev. A} \textbf{\bibinfo{volume}{102}},
  \bibinfo{pages}{063107} (\bibinfo{year}{2020}).

\bibitem[{\citenamefont{Koll{\'a}r et~al.}(2019)\citenamefont{Koll{\'a}r,
  Fitzpatrick, and Houck}}]{Kollar_2019}
\bibinfo{author}{\bibfnamefont{A.~J.} \bibnamefont{Koll{\'a}r}},
  \bibinfo{author}{\bibfnamefont{M.}~\bibnamefont{Fitzpatrick}},
  \bibnamefont{and} \bibinfo{author}{\bibfnamefont{A.~A.} \bibnamefont{Houck}},
  \bibinfo{journal}{Nature} \textbf{\bibinfo{volume}{571}}, \bibinfo{pages}{45}
  (\bibinfo{year}{2019}).

\bibitem[{\citenamefont{Carusotto et~al.}(2020)\citenamefont{Carusotto, Houck,
  Koll{\'a}r, Roushan, Schuster, and Simon}}]{Carusotto_2020}
\bibinfo{author}{\bibfnamefont{I.}~\bibnamefont{Carusotto}},
  \bibinfo{author}{\bibfnamefont{A.~A.} \bibnamefont{Houck}},
  \bibinfo{author}{\bibfnamefont{A.~J.} \bibnamefont{Koll{\'a}r}},
  \bibinfo{author}{\bibfnamefont{P.}~\bibnamefont{Roushan}},
  \bibinfo{author}{\bibfnamefont{D.~I.} \bibnamefont{Schuster}},
  \bibnamefont{and} \bibinfo{author}{\bibfnamefont{J.}~\bibnamefont{Simon}},
  \bibinfo{journal}{Nature Phys.} \textbf{\bibinfo{volume}{16}},
  \bibinfo{pages}{268} (\bibinfo{year}{2020}).

\bibitem[{\citenamefont{Jo et~al.}(2012)\citenamefont{Jo, Guzman, Thomas,
  Hosur, Vishwanath, and Stamper-Kurn}}]{Jo_2012}
\bibinfo{author}{\bibfnamefont{G.-B.} \bibnamefont{Jo}},
  \bibinfo{author}{\bibfnamefont{J.}~\bibnamefont{Guzman}},
  \bibinfo{author}{\bibfnamefont{C.~K.} \bibnamefont{Thomas}},
  \bibinfo{author}{\bibfnamefont{P.}~\bibnamefont{Hosur}},
  \bibinfo{author}{\bibfnamefont{A.}~\bibnamefont{Vishwanath}},
  \bibnamefont{and} \bibinfo{author}{\bibfnamefont{D.~M.}
  \bibnamefont{Stamper-Kurn}}, \bibinfo{journal}{Phys. Rev. Lett.}
  \textbf{\bibinfo{volume}{108}}, \bibinfo{pages}{045305}
  (\bibinfo{year}{2012}).

\bibitem[{\citenamefont{Sbroscia et~al.}(2020)\citenamefont{Sbroscia, Viebahn,
  Carter, Yu, Gaunt, and Schneider}}]{Sbroscia_2020}
\bibinfo{author}{\bibfnamefont{M.}~\bibnamefont{Sbroscia}},
  \bibinfo{author}{\bibfnamefont{K.}~\bibnamefont{Viebahn}},
  \bibinfo{author}{\bibfnamefont{E.}~\bibnamefont{Carter}},
  \bibinfo{author}{\bibfnamefont{J.-C.} \bibnamefont{Yu}},
  \bibinfo{author}{\bibfnamefont{A.}~\bibnamefont{Gaunt}}, \bibnamefont{and}
  \bibinfo{author}{\bibfnamefont{U.}~\bibnamefont{Schneider}},
  \bibinfo{journal}{Phys. Rev. Lett.} \textbf{\bibinfo{volume}{125}},
  \bibinfo{pages}{200604} (\bibinfo{year}{2020}).

\bibitem[{\citenamefont{Freedman et~al.}(2006)\citenamefont{Freedman, Bartal,
  Segev, Lifshitz, Christodoulides, and Fleischer}}]{Freedman_2006}
\bibinfo{author}{\bibfnamefont{B.}~\bibnamefont{Freedman}},
  \bibinfo{author}{\bibfnamefont{G.}~\bibnamefont{Bartal}},
  \bibinfo{author}{\bibfnamefont{M.}~\bibnamefont{Segev}},
  \bibinfo{author}{\bibfnamefont{R.}~\bibnamefont{Lifshitz}},
  \bibinfo{author}{\bibfnamefont{D.~N.} \bibnamefont{Christodoulides}},
  \bibnamefont{and} \bibinfo{author}{\bibfnamefont{J.~W.}
  \bibnamefont{Fleischer}}, \bibinfo{journal}{Nature}
  \textbf{\bibinfo{volume}{440}}, \bibinfo{pages}{1166} (\bibinfo{year}{2006}).

\bibitem[{\citenamefont{Baboux et~al.}(2017)\citenamefont{Baboux, Levy,
  Lema{\^\i}tre, G{\'o}mez, Galopin, Le~Gratiet, Sagnes, Amo, Bloch, and
  Akkermans}}]{Baboux_2017}
\bibinfo{author}{\bibfnamefont{F.}~\bibnamefont{Baboux}},
  \bibinfo{author}{\bibfnamefont{E.}~\bibnamefont{Levy}},
  \bibinfo{author}{\bibfnamefont{A.}~\bibnamefont{Lema{\^\i}tre}},
  \bibinfo{author}{\bibfnamefont{C.}~\bibnamefont{G{\'o}mez}},
  \bibinfo{author}{\bibfnamefont{E.}~\bibnamefont{Galopin}},
  \bibinfo{author}{\bibfnamefont{L.}~\bibnamefont{Le~Gratiet}},
  \bibinfo{author}{\bibfnamefont{I.}~\bibnamefont{Sagnes}},
  \bibinfo{author}{\bibfnamefont{A.}~\bibnamefont{Amo}},
  \bibinfo{author}{\bibfnamefont{J.}~\bibnamefont{Bloch}}, \bibnamefont{and}
  \bibinfo{author}{\bibfnamefont{E.}~\bibnamefont{Akkermans}},
  \bibinfo{journal}{Phys. Rev. B} \textbf{\bibinfo{volume}{95}},
  \bibinfo{pages}{161114(R)} (\bibinfo{year}{2017}).

\bibitem[{\citenamefont{Kempkes et~al.}(2018)\citenamefont{Kempkes, Slot,
  Freeney, Zevenhuizen, Vanmaekelbergh, Swart, and Smith}}]{Kempkes_2018}
\bibinfo{author}{\bibfnamefont{S.~N.} \bibnamefont{Kempkes}},
  \bibinfo{author}{\bibfnamefont{M.~R.} \bibnamefont{Slot}},
  \bibinfo{author}{\bibfnamefont{S.~E.} \bibnamefont{Freeney}},
  \bibinfo{author}{\bibfnamefont{S.~J.~M.} \bibnamefont{Zevenhuizen}},
  \bibinfo{author}{\bibfnamefont{D.}~\bibnamefont{Vanmaekelbergh}},
  \bibinfo{author}{\bibfnamefont{I.}~\bibnamefont{Swart}}, \bibnamefont{and}
  \bibinfo{author}{\bibfnamefont{C.~M.} \bibnamefont{Smith}},
  \bibinfo{journal}{Nature Phys.} \textbf{\bibinfo{volume}{15}},
  \bibinfo{pages}{127} (\bibinfo{year}{2018}).

\bibitem[{\citenamefont{Periwal et~al.}(2021)\citenamefont{Periwal, Cooper,
  Kunkel, Wienand, Davis, and Schleier-Smith}}]{Periwal_2021}
\bibinfo{author}{\bibfnamefont{A.}~\bibnamefont{Periwal}},
  \bibinfo{author}{\bibfnamefont{E.~S.} \bibnamefont{Cooper}},
  \bibinfo{author}{\bibfnamefont{P.}~\bibnamefont{Kunkel}},
  \bibinfo{author}{\bibfnamefont{J.~F.} \bibnamefont{Wienand}},
  \bibinfo{author}{\bibfnamefont{E.~J.} \bibnamefont{Davis}}, \bibnamefont{and}
  \bibinfo{author}{\bibfnamefont{M.}~\bibnamefont{Schleier-Smith}},
  \bibinfo{journal}{Nature} \textbf{\bibinfo{volume}{600}},
  \bibinfo{pages}{630–635} (\bibinfo{year}{2021}).

\bibitem[{\citenamefont{Burioni et~al.}(2000)\citenamefont{Burioni, Cassi,
  Meccoli, Rasetti, Regina, Sodano, and Vezzani}}]{Burioni_2000}
\bibinfo{author}{\bibfnamefont{R.}~\bibnamefont{Burioni}},
  \bibinfo{author}{\bibfnamefont{D.}~\bibnamefont{Cassi}},
  \bibinfo{author}{\bibfnamefont{I.}~\bibnamefont{Meccoli}},
  \bibinfo{author}{\bibfnamefont{M.}~\bibnamefont{Rasetti}},
  \bibinfo{author}{\bibfnamefont{S.}~\bibnamefont{Regina}},
  \bibinfo{author}{\bibfnamefont{P.}~\bibnamefont{Sodano}}, \bibnamefont{and}
  \bibinfo{author}{\bibfnamefont{A.}~\bibnamefont{Vezzani}},
  \bibinfo{journal}{EPL} \textbf{\bibinfo{volume}{52}}, \bibinfo{pages}{251}
  (\bibinfo{year}{2000}).

\bibitem[{\citenamefont{Buonsante et~al.}(2004)\citenamefont{Buonsante, Penna,
  and Vezzani}}]{Buonsante_2004}
\bibinfo{author}{\bibfnamefont{P.}~\bibnamefont{Buonsante}},
  \bibinfo{author}{\bibfnamefont{V.}~\bibnamefont{Penna}}, \bibnamefont{and}
  \bibinfo{author}{\bibfnamefont{A.}~\bibnamefont{Vezzani}},
  \bibinfo{journal}{Phys. Rev. B} \textbf{\bibinfo{volume}{70}},
  \bibinfo{pages}{184520} (\bibinfo{year}{2004}).

\bibitem[{\citenamefont{Sodano et~al.}(2006)\citenamefont{Sodano, Trombettoni,
  Silvestrini, Russo, and Ruggiero}}]{Sodano_2006}
\bibinfo{author}{\bibfnamefont{P.}~\bibnamefont{Sodano}},
  \bibinfo{author}{\bibfnamefont{A.}~\bibnamefont{Trombettoni}},
  \bibinfo{author}{\bibfnamefont{P.}~\bibnamefont{Silvestrini}},
  \bibinfo{author}{\bibfnamefont{R.}~\bibnamefont{Russo}}, \bibnamefont{and}
  \bibinfo{author}{\bibfnamefont{B.}~\bibnamefont{Ruggiero}},
  \bibinfo{journal}{New J. Phys.} \textbf{\bibinfo{volume}{8}},
  \bibinfo{pages}{327} (\bibinfo{year}{2006}).

\bibitem[{\citenamefont{Tonks}(1936)}]{Tonks_1936}
\bibinfo{author}{\bibfnamefont{L.}~\bibnamefont{Tonks}},
  \bibinfo{journal}{Phys. Rev.} \textbf{\bibinfo{volume}{50}},
  \bibinfo{pages}{955} (\bibinfo{year}{1936}).

\bibitem[{\citenamefont{Girardeau}(1960)}]{Girardeau_1960}
\bibinfo{author}{\bibfnamefont{M.}~\bibnamefont{Girardeau}},
  \bibinfo{journal}{J. Math. Phys.} \textbf{\bibinfo{volume}{1}},
  \bibinfo{pages}{516} (\bibinfo{year}{1960}).

\bibitem[{\citenamefont{Arovas et~al.}(1985)\citenamefont{Arovas, Schrieffer,
  Wilczek, and Zee}}]{Arovas_1985}
\bibinfo{author}{\bibfnamefont{D.~P.} \bibnamefont{Arovas}},
  \bibinfo{author}{\bibfnamefont{R.}~\bibnamefont{Schrieffer}},
  \bibinfo{author}{\bibfnamefont{F.}~\bibnamefont{Wilczek}}, \bibnamefont{and}
  \bibinfo{author}{\bibfnamefont{A.}~\bibnamefont{Zee}}, \bibinfo{journal}{Nuc.
  Phys. B} \textbf{\bibinfo{volume}{251}}, \bibinfo{pages}{117}
  (\bibinfo{year}{1985}).

\bibitem[{\citenamefont{Sobirov et~al.}(2010)\citenamefont{Sobirov, Matrasulov,
  Sabirov, Sawada, and Nakamura}}]{Sobirov_2010}
\bibinfo{author}{\bibfnamefont{Z.}~\bibnamefont{Sobirov}},
  \bibinfo{author}{\bibfnamefont{D.}~\bibnamefont{Matrasulov}},
  \bibinfo{author}{\bibfnamefont{K.}~\bibnamefont{Sabirov}},
  \bibinfo{author}{\bibfnamefont{S.-i.} \bibnamefont{Sawada}},
  \bibnamefont{and} \bibinfo{author}{\bibfnamefont{K.}~\bibnamefont{Nakamura}},
  \bibinfo{journal}{Phys. Rev. E} \textbf{\bibinfo{volume}{81}},
  \bibinfo{pages}{066602} (\bibinfo{year}{2010}).

\bibitem[{\citenamefont{Diep et~al.}(2013)}]{Diep_Book}
\bibinfo{author}{\bibfnamefont{H.}~\bibnamefont{Diep}} \bibnamefont{et~al.},
  \emph{\bibinfo{title}{Frustrated spin systems}} (\bibinfo{publisher}{World
  scientific}, \bibinfo{year}{2013}).

\bibitem[{\citenamefont{C\'espedes}(2018)}]{Cespedes_2018}
\bibinfo{author}{\bibfnamefont{P.}~\bibnamefont{C\'espedes}}, Master's thesis,
  \bibinfo{school}{FAMAF, Universidad Nacional de C\'ordoba}
  (\bibinfo{year}{2018}).

\bibitem[{\citenamefont{Combescot et~al.}(2008)\citenamefont{Combescot,
  Betbeder-Matibet, and Dubin}}]{combescot_2008}
\bibinfo{author}{\bibfnamefont{M.}~\bibnamefont{Combescot}},
  \bibinfo{author}{\bibfnamefont{O.}~\bibnamefont{Betbeder-Matibet}},
  \bibnamefont{and} \bibinfo{author}{\bibfnamefont{F.}~\bibnamefont{Dubin}},
  \bibinfo{journal}{Phys. Rep.} \textbf{\bibinfo{volume}{463}},
  \bibinfo{pages}{215} (\bibinfo{year}{2008}).

\bibitem[{\citenamefont{Tennie et~al.}(2017)\citenamefont{Tennie, Vedral, and
  Schilling}}]{Tennie_2017}
\bibinfo{author}{\bibfnamefont{F.}~\bibnamefont{Tennie}},
  \bibinfo{author}{\bibfnamefont{V.}~\bibnamefont{Vedral}}, \bibnamefont{and}
  \bibinfo{author}{\bibfnamefont{C.}~\bibnamefont{Schilling}},
  \bibinfo{journal}{Phys. Rev. B} \textbf{\bibinfo{volume}{96}},
  \bibinfo{pages}{064502} (\bibinfo{year}{2017}).

\bibitem[{\citenamefont{Tichy et~al.}(2012{\natexlab{a}})\citenamefont{Tichy,
  Bouvrie, and M\o{}lmer}}]{tichy_2012b}
\bibinfo{author}{\bibfnamefont{M.~C.} \bibnamefont{Tichy}},
  \bibinfo{author}{\bibfnamefont{P.~A.} \bibnamefont{Bouvrie}},
  \bibnamefont{and}
  \bibinfo{author}{\bibfnamefont{K.}~\bibnamefont{M\o{}lmer}},
  \bibinfo{journal}{Phys. Rev. A} \textbf{\bibinfo{volume}{86}},
  \bibinfo{pages}{042317} (\bibinfo{year}{2012}{\natexlab{a}}).

\bibitem[{\citenamefont{C\'espedes et~al.}(2019)\citenamefont{C\'espedes,
  Rufeil-Fiori, Bouvrie, Majtey, and Cormick}}]{Cespedes_2019}
\bibinfo{author}{\bibfnamefont{P.}~\bibnamefont{C\'espedes}},
  \bibinfo{author}{\bibfnamefont{E.}~\bibnamefont{Rufeil-Fiori}},
  \bibinfo{author}{\bibfnamefont{P.~A.} \bibnamefont{Bouvrie}},
  \bibinfo{author}{\bibfnamefont{A.~P.} \bibnamefont{Majtey}},
  \bibnamefont{and} \bibinfo{author}{\bibfnamefont{C.}~\bibnamefont{Cormick}},
  \bibinfo{journal}{Phys. Rev. A} \textbf{\bibinfo{volume}{100}},
  \bibinfo{pages}{012309} (\bibinfo{year}{2019}).

\bibitem[{\citenamefont{Cuestas and Cormick}(2022)}]{Cuestas_Cormick_2022}
\bibinfo{author}{\bibfnamefont{E.}~\bibnamefont{Cuestas}} \bibnamefont{and}
  \bibinfo{author}{\bibfnamefont{C.}~\bibnamefont{Cormick}},
  \bibinfo{journal}{Phys. Rev. A} \textbf{\bibinfo{volume}{105}},
  \bibinfo{pages}{013302} (\bibinfo{year}{2022}).

\bibitem[{\citenamefont{Bernardet et~al.}(2002)\citenamefont{Bernardet,
  Batrouni, Meunier, Schmid, Troyer, and Dorneich}}]{Bernardet_2002}
\bibinfo{author}{\bibfnamefont{K.}~\bibnamefont{Bernardet}},
  \bibinfo{author}{\bibfnamefont{G.}~\bibnamefont{Batrouni}},
  \bibinfo{author}{\bibfnamefont{J.-L.} \bibnamefont{Meunier}},
  \bibinfo{author}{\bibfnamefont{G.}~\bibnamefont{Schmid}},
  \bibinfo{author}{\bibfnamefont{M.}~\bibnamefont{Troyer}}, \bibnamefont{and}
  \bibinfo{author}{\bibfnamefont{A.}~\bibnamefont{Dorneich}},
  \bibinfo{journal}{Phys. Rev. B} \textbf{\bibinfo{volume}{65}},
  \bibinfo{pages}{104519} (\bibinfo{year}{2002}).

\bibitem[{\citenamefont{Falconer}(2003)}]{Falconer_2003}
\bibinfo{author}{\bibfnamefont{K.}~\bibnamefont{Falconer}},
  \emph{\bibinfo{title}{Fractal Geometry}} (\bibinfo{publisher}{John Wiley \&
  Sons, Ltd}, \bibinfo{year}{2003}), ISBN \bibinfo{isbn}{9780470013854}.

\bibitem[{\citenamefont{Baxter}(1982)}]{baxter_book}
\bibinfo{author}{\bibfnamefont{R.~J.} \bibnamefont{Baxter}},
  \emph{\bibinfo{title}{Exactly solved models in statistical mechanics}}
  (\bibinfo{publisher}{Academic Press, London}, \bibinfo{year}{1982}).

\bibitem[{\citenamefont{Bethe}(1931)}]{bethe_1931}
\bibinfo{author}{\bibfnamefont{H.}~\bibnamefont{Bethe}}, \bibinfo{journal}{Z.
  Phys. A}  (\bibinfo{year}{1931}).

\bibitem[{\citenamefont{Karabach et~al.}(1997)\citenamefont{Karabach, Muller,
  Gould, and Tobochnik}}]{karabach_1997}
\bibinfo{author}{\bibfnamefont{M.}~\bibnamefont{Karabach}},
  \bibinfo{author}{\bibfnamefont{G.}~\bibnamefont{Muller}},
  \bibinfo{author}{\bibfnamefont{H.}~\bibnamefont{Gould}}, \bibnamefont{and}
  \bibinfo{author}{\bibfnamefont{J.}~\bibnamefont{Tobochnik}},
  \bibinfo{journal}{Comp. in Phys.} \textbf{\bibinfo{volume}{11}},
  \bibinfo{pages}{36} (\bibinfo{year}{1997}).

\bibitem[{\citenamefont{Blundell}(2001)}]{Blundell_book}
\bibinfo{author}{\bibfnamefont{S.}~\bibnamefont{Blundell}},
  \emph{\bibinfo{title}{Magnetism in Condensed Matter}}
  (\bibinfo{publisher}{Oxford Master Series in Cond. Matt. Phys.},
  \bibinfo{year}{2001}).

\bibitem[{\citenamefont{Tomczak et~al.}(1996)\citenamefont{Tomczak, Ferchmin,
  and Richter}}]{Tomczak_1996}
\bibinfo{author}{\bibfnamefont{P.}~\bibnamefont{Tomczak}},
  \bibinfo{author}{\bibfnamefont{A.}~\bibnamefont{Ferchmin}}, \bibnamefont{and}
  \bibinfo{author}{\bibfnamefont{J.}~\bibnamefont{Richter}},
  \bibinfo{journal}{Phys. Rev. B} \textbf{\bibinfo{volume}{54}},
  \bibinfo{pages}{395} (\bibinfo{year}{1996}).

\bibitem[{\citenamefont{Voigt et~al.}(1998)\citenamefont{Voigt, Richter, and
  Tomczak}}]{Voigt_1998}
\bibinfo{author}{\bibfnamefont{A.}~\bibnamefont{Voigt}},
  \bibinfo{author}{\bibfnamefont{J.}~\bibnamefont{Richter}}, \bibnamefont{and}
  \bibinfo{author}{\bibfnamefont{P.}~\bibnamefont{Tomczak}},
  \bibinfo{journal}{J. Magn. Magn. Mater.} \textbf{\bibinfo{volume}{183}},
  \bibinfo{pages}{68} (\bibinfo{year}{1998}).

\bibitem[{\citenamefont{Voigt et~al.}(2001)\citenamefont{Voigt, Richter, and
  Tomczak}}]{Voigt_2001}
\bibinfo{author}{\bibfnamefont{A.}~\bibnamefont{Voigt}},
  \bibinfo{author}{\bibfnamefont{J.}~\bibnamefont{Richter}}, \bibnamefont{and}
  \bibinfo{author}{\bibfnamefont{P.}~\bibnamefont{Tomczak}},
  \bibinfo{journal}{Phys. A: Stat. Mech. Appl.} \textbf{\bibinfo{volume}{299}},
  \bibinfo{pages}{461} (\bibinfo{year}{2001}).

\bibitem[{\citenamefont{Tichy et~al.}(2012{\natexlab{b}})\citenamefont{Tichy,
  Bouvrie, and M{\o}lmer}}]{tichy_2012}
\bibinfo{author}{\bibfnamefont{M.~C.} \bibnamefont{Tichy}},
  \bibinfo{author}{\bibfnamefont{P.~A.} \bibnamefont{Bouvrie}},
  \bibnamefont{and}
  \bibinfo{author}{\bibfnamefont{K.}~\bibnamefont{M{\o}lmer}},
  \bibinfo{journal}{Phys. Rev. Lett.} \textbf{\bibinfo{volume}{109}},
  \bibinfo{pages}{260403} (\bibinfo{year}{2012}{\natexlab{b}}).

\bibitem[{\citenamefont{Lasmar et~al.}(2019)\citenamefont{Lasmar, Bouvrie,
  Sajna, Tichy, and Kurzy{\'n}ski}}]{lasmar_2019}
\bibinfo{author}{\bibfnamefont{Z.}~\bibnamefont{Lasmar}},
  \bibinfo{author}{\bibfnamefont{P.~A.} \bibnamefont{Bouvrie}},
  \bibinfo{author}{\bibfnamefont{A.~S.} \bibnamefont{Sajna}},
  \bibinfo{author}{\bibfnamefont{M.~C.} \bibnamefont{Tichy}}, \bibnamefont{and}
  \bibinfo{author}{\bibfnamefont{P.}~\bibnamefont{Kurzy{\'n}ski}},
  \bibinfo{journal}{Phys. Rev. A} \textbf{\bibinfo{volume}{100}},
  \bibinfo{pages}{032105} (\bibinfo{year}{2019}).

\bibitem[{\citenamefont{Law}(2005)}]{law_2005}
\bibinfo{author}{\bibfnamefont{C.~K.} \bibnamefont{Law}},
  \bibinfo{journal}{Phys. Rev. A} \textbf{\bibinfo{volume}{71}},
  \bibinfo{pages}{034306} (\bibinfo{year}{2005}).

\bibitem[{\citenamefont{{Combescot, M.} et~al.}(2003)\citenamefont{{Combescot,
  M.}, {Leyronas, X.}, and {Tanguy, C.}}}]{combescot_2003}
\bibinfo{author}{\bibnamefont{{Combescot, M.}}},
  \bibinfo{author}{\bibnamefont{{Leyronas, X.}}}, \bibnamefont{and}
  \bibinfo{author}{\bibnamefont{{Tanguy, C.}}}, \bibinfo{journal}{Eur. Phys. J.
  B} \textbf{\bibinfo{volume}{31}}, \bibinfo{pages}{17} (\bibinfo{year}{2003}).

\bibitem[{\citenamefont{Chudzicki et~al.}(2010)\citenamefont{Chudzicki, Oke,
  and Wootters}}]{chudzicki_2010}
\bibinfo{author}{\bibfnamefont{C.}~\bibnamefont{Chudzicki}},
  \bibinfo{author}{\bibfnamefont{O.}~\bibnamefont{Oke}}, \bibnamefont{and}
  \bibinfo{author}{\bibfnamefont{W.~K.} \bibnamefont{Wootters}},
  \bibinfo{journal}{Phys. Rev. Lett.} \textbf{\bibinfo{volume}{104}},
  \bibinfo{pages}{070402} (\bibinfo{year}{2010}).

\bibitem[{\citenamefont{Combescot}(2011)}]{combescot_2011b}
\bibinfo{author}{\bibfnamefont{M.}~\bibnamefont{Combescot}},
  \bibinfo{journal}{EPL} \textbf{\bibinfo{volume}{96}}, \bibinfo{pages}{60002}
  (\bibinfo{year}{2011}).

\bibitem[{\citenamefont{Combescot and Shiau}(2015)}]{combescot_2015}
\bibinfo{author}{\bibfnamefont{M.}~\bibnamefont{Combescot}} \bibnamefont{and}
  \bibinfo{author}{\bibfnamefont{S.-Y.} \bibnamefont{Shiau}},
  \emph{\bibinfo{title}{Excitons and Cooper Pairs: Two Composite Bosons in
  Many-Body Physics}} (\bibinfo{publisher}{Oxford University Press},
  \bibinfo{year}{2015}).

\bibitem[{\citenamefont{Jiménez et~al.}(2023)\citenamefont{Jiménez, Cuestas,
  Majtey, and Cormick}}]{Jimenez_2023}
\bibinfo{author}{\bibfnamefont{M.~D.} \bibnamefont{Jiménez}},
  \bibinfo{author}{\bibfnamefont{E.}~\bibnamefont{Cuestas}},
  \bibinfo{author}{\bibfnamefont{A.~P.} \bibnamefont{Majtey}},
  \bibnamefont{and} \bibinfo{author}{\bibfnamefont{C.}~\bibnamefont{Cormick}},
  \bibinfo{journal}{SciPost Phys. Core} \textbf{\bibinfo{volume}{6}},
  \bibinfo{pages}{012} (\bibinfo{year}{2023}).

\bibitem[{\citenamefont{Vicsek}(1983)}]{Vicsek_1983}
\bibinfo{author}{\bibfnamefont{T.}~\bibnamefont{Vicsek}}, \bibinfo{journal}{J.
  Phys. A: Math. Gen.} \textbf{\bibinfo{volume}{16}}, \bibinfo{pages}{L647}
  (\bibinfo{year}{1983}).

\bibitem[{\citenamefont{Blumen et~al.}(2004)\citenamefont{Blumen, von Ferber,
  Jurjiu, and Koslowski}}]{Blumen_2004}
\bibinfo{author}{\bibfnamefont{A.}~\bibnamefont{Blumen}},
  \bibinfo{author}{\bibfnamefont{C.}~\bibnamefont{von Ferber}},
  \bibinfo{author}{\bibfnamefont{A.}~\bibnamefont{Jurjiu}}, \bibnamefont{and}
  \bibinfo{author}{\bibfnamefont{T.}~\bibnamefont{Koslowski}},
  \bibinfo{journal}{Macromolecules} \textbf{\bibinfo{volume}{37}},
  \bibinfo{pages}{638} (\bibinfo{year}{2004}).

\bibitem[{\citenamefont{Berge}(2001)}]{Berge_2001}
\bibinfo{author}{\bibfnamefont{C.}~\bibnamefont{Berge}},
  \emph{\bibinfo{title}{The theory of graphs}} (\bibinfo{publisher}{Courier
  Corporation}, \bibinfo{year}{2001}).

\bibitem[{\citenamefont{Newman}(2018)}]{newman2018networks}
\bibinfo{author}{\bibfnamefont{M.}~\bibnamefont{Newman}},
  \emph{\bibinfo{title}{Networks}} (\bibinfo{publisher}{OUP Oxford},
  \bibinfo{year}{2018}), ISBN \bibinfo{isbn}{9780192527493}.

\end{thebibliography}
\end{document}